\documentclass[aps,preprint]{revtex4-1}
\usepackage{graphicx}
\usepackage{amssymb}
\usepackage{amsmath}
\usepackage{color}
\usepackage{comment}
\usepackage{enumerate}
\usepackage{bm}
\begin{document}
%\title {Dipolar Interactions in Magnetic Fluids: Consequences and Applications}
\title {Non-Uniform Hysteresis in Small Clusters of Magnetic Nanoparticles}
 
\author{Manish Anand}
\email{itsanand121@gmail.com}
\affiliation{Department of Physics, Bihar National College, Patna University, Patna-800004, India.}

\date{\today}
\begin{abstract}
Using first-principle calculations and kinetic Monte Carlo simulation, we study the local and averaged hysteresis in tiny clusters of $k$ magnetic nanoparticles (MNPs) or $k$-mers. 
%To understand the effect of dipolar interaction at the particle level, 
We also analyze the variation of local dipolar field acting on the constituent nanoparticles as a function of the external magnetic field. The dipolar interaction is found to promote chain-like arrangement in such a cluster. Irrespective of cluster size, the local hysteresis response depends strongly on the corresponding dipolar field acted on a nanoparticle. In a small $k$-mer, there is a wide variation in local hysteresis as a function of nanoparticle position. On the other hand, the local hysteresis is more uniform for larger $k$-mer, except for MNPs at the boundary. In the case of superparamagnetic nanoparticle and weak dipolar interaction, the local hysteresis loop area $A^{}_i$ is minimal and depends weakly on the  $k$-mer size. While for ferromagnetic counterpart, $A^{}_i$ is considerably large even for weakly interacting MNPs. The value of $A^{}_i$ is found to be directly proportional to the dipolar field acting on the nanoparticle. 
%The coercivity and remanence also get enhanced with an increase in dipolar interaction strength and size of $k$-mer. 
The dipolar interaction and $k$-mer size also enhances the coercivity and remanence. There is always an increase in $A^{}_i$ with clutser size and dipolar interaction strength. 
Similarly, the averaged hysteresis loop area $A$ also depends strongly on the $k$-mer size, particle size and dipolar interaction strength. $A$ and $A^{}_i$ always increase with $k$-mer size and  dipolar interaction strength. 
Interestingly, the value of $A$ saturates for $k\geq20$ and considerable dipolar interaction irrespective of particle size. 
%These results should be kept in mind for usage of MNPs in drug delivery and hyperthermia applications where clustering is prominent. 
We believe that the present work would help understand the intricate role of dipolar interaction on hysteresis and organizational structure of MNPs and their usage in drug delivery and hyperthermia applications. 	 
\end{abstract}
%\pacs{47.65.Cb, 47.57.eb, 75.60.-d, 87.50.wp}
\maketitle
\section{Introduction}
\label{1}
%Magnetic fluids (MFs) are colloidal dispersions of magnetic nanopraticles (MNPs). They have been a subject of great interest ever since their discovery in the early 1960s \cite{pankhurst2003,purushotham2010,gijs2010,odenbach2002,pankhurst2009,mehdaoui2013,gubin2009,rosensweig1997}. Like any other colloidal suspension, MNPs may form large aggregates that eventually flocculate and destroy the stability of the magnetic fluid.  To prevent this, they are generally coated by a neutral surfactant. The latter stabilises the  suspension by providing steric hinderance which counteracts the dipolar attraction amongst the MNPs. Consequently, they are dispersed in the suspension with randomly oriented magnetic moments. Such particles exhibit N\'{e}el and Brownian relaxation which determine the response of the MF to applied fields. The relaxation times are strongly size-dependent, and are also greatly influenced by material parameters such as magnetic moment, anisotropy, etc \cite{rosensweig1997,w1996,singh2009,eberbeck2006}.  %These can be easily tailored to suit applications.
%The last decade has revealed the tremendous potential of magnetic nanoparticles (MNPs) for diagnostic and therapeutic applications
Magnetic nanoparticles (MNPs) have tremendous potential in diagnostic and therapeutic applications~\cite{mitchell2020,yu2020,bustamante2013,wang2012,schutz2014,anand2016}. The reasons behind their applicability are many~\cite{smith2017,huang2016}. 
%They can be easily detected and manipulated by the application of external magnetic fields. 
For instance, they can be bound to drugs, dyes and biological entities, thereby providing opportunities for site-specific drug delivery, improved quality of magnetic resonance imaging, manipulation of cell membranes, etc.~\cite{li2020,cheng2013,huang2012}. Most importantly, they can be made to dissipate heat when subjected to an oscillating magnetic field. Therefore, when targeted on malignant tumours, MNPs can kill the cancerous cells or introduce a modest rise in temperature to increase the efficacy of drug reactions with the body. This phenomenon results in selective warming of the local area and is usually referred to as hyperthermia in the medical literature~\cite{mehdaoui2013,lee2011,pankhurst2003}.         

Several procedures, often ad-hoc, are being used for hyperthermia. Usually, dilute solutions of MNPs (coated, functionalized and often bound to chemotherapeutic drugs) are injected into the bloodstream and magnetically targeted to affected tissues or organs~\cite{obaidat2019,shubitidze2015}. An advantage here is their rapid clearance from the body by way of renal and biliary excretions. 
%Further, the low concentrations reduce the risks of aggregation and embolism.
A significant disadvantage, however, is the danger of damage to healthy cells. 
There is also the possibility of insufficient heating and inefficient drug delivery due to particle dispersion. There has been a surge of activity in recent years to design {\it temperature-sensitive liposomes} to overcome this shortcoming~\cite {gao2019,chen2020,allen2013,maeki2018,allen2004}. The latter are lipid sacs encapsulating fluids containing chemotherapeutic drugs and MNPs. The membranes are chosen to be biocompatible, with a melting temperature slightly above the body temperature. On applying an oscillating magnetic field, the contents are released when the heat generated by the enclosed MNPs is sufficient to melt the lipid membrane. 
%The MNPs continue to be useful, now as heaters for the malignant cells. 
Therefore, these micron-sized ferries hold tremendous promise, as they combine the advantages of targeted drug delivery with localized hyperthermia treatment. 

Many theoretical descriptions have emerged to understand heat dissipation in such a system. These are generally based on single-particle models which ignore the inter-particle dipole-dipole interactions~\cite{usov2010,mehdaoui2011,rosensweig2002}.  However, wide discrepancies between experimental measurements of heat dissipation and corresponding theoretical calculations have been reported in the literature~\cite{nemati2018,haase2012}. Investigations using sophisticated experimental techniques such as  electron magnetic resonance (EMR), transmission electron microscopy (TEM), zero-field birefringence, etc. have indicated small agglomerates of MNPs in these carriers~\cite{hu2014,osterloh2005}. It is not unexpected, recalling the presence of the ubiquitous dipolar interaction~\cite{anand2020,odenbach2002,rosensweig1997}. The latter is long-ranged and anisotropic, which has a diverse effect on the morphologies and magnetic properties of an assembly of MNPs~\cite{branquinho2013,anand12020,anand2021}.
%{\color{blue}Heat dissipation is a consequence of the delayed response of the magnetic moment to an applied oscillating magnetic field. It is therefore expected to critically depend upon the relaxation times of the clusters. These are significantly different from single particle values due to a modification of both, the hydrodynamic radius and magnetic volume.} 
Further, the constituting MNPs may form clusters in many different ways. Therefore, there is a plethora of geometric configurations and magnetic moments orientations.
%, each  characterised by a distinct  {\it effective} hydrodynamic radius and magnetic volume. The distribution of relaxation times of these {\it nano heaters} and consequently the heat dissipated by them is unexpectedly wide spread. 
Consequently, it is crucial to understand the role of dipolar interactions, as a function of cluster's geometry and the orientation of magnetic moments, to identify the efficient heat generators from the inefficient ones to optimise heat dissipation for their efficient usage in hyperthermia applications.

%The dipolar interaction is long-ranged and anisotropic in nature. 
In MNPs assembly, the impact of dipolar interaction should not be overlooked as clustering of nanoparticles is also caused by the cellular environment and other factors such as the spatial variation in the applied magnetic field, nanoparticle-synthesis procedures, etc.~\cite{vikesland2016,ye2016}. As a consequence, the dipolar interaction drastically affects the amount of heat dissipation in such assembly. Therefore, the study of magnetic hysteresis as a function of dipolar interaction is equally important for better usage of MNPs in hyperthermia applications. Generally, such a system's heating efficiency is measured by the specific absorption ratio (SAR), which is related to the hysteresis loop area and the frequency of the applied alternating magnetic field~\cite{carrey2011}. On the other hand, some recent works stress the importance of local heat dissipation by the constituent nanoparticles~\cite{munoz2020,torche2020}. Recent studies also suggest that the nanoparticle's local heat is way more useful than the entire system's averaged heating~\cite{lin2020,cazares2019,riedinger2013,dias2013}. 
Therefore, it is crucial to understand the role of dipolar interaction on the local (individual nanoparticle) and the averaged hysteresis response of the underlying system. To probe it more microscopically,  it is also valuable to understand the variation of local dipolar field acted on each nanoparticle as a function of the applied magnetic field.

This paper aims to understand the consequences of dipolar interactions and their implications on the organizational structure of small cluster of nanoparticles; local (due to the individual nanoparticle) and averaged hysteresis response of such a system. 
%We also probe the hysteresis response of constituting particle in such a system. 
We also study the detailed mechanism of the variation of local dipolar field acting on the constituent nanoparticle as a function of the external magnetic field. We first identify the low energy configurations of clusters of $k$ MNPs or $k$-mers  ($k=2$, 3, 4, 5, etc.) by a first principle calculation. We then use the kinetic Monte Carlo (kMC) simulation technique to study the local and the global hysteresis behaviour as a function of dipolar interaction strength, size of $k$-mer and particle size. To probe further, we also perform a detailed analysis of the local dipolar field acted on each nanoparticle by all other MNPs present in the system as a function of the external magnetic field.

The rest of the paper is organized as follows. In Section II, we discuss the methodologies required to evaluate the low-energy configurations. kMC algorithm is also discussed in brief.
%calculate relaxation times of clusters, heat dissipation and corresponding temperature rise are provided.
In Section III, we present the numerical results. 
%use the developed framework to evaluate relaxation times and heat dissipation in the commonly used dilute suspension of magnetite ($Fe_{3}O_{4}$) nanoparticles. A practical application of our results for hyperthermia using (i) MFs  and (ii) liposomes for targeted therapeutics is discussed in Section IV. 
Finally, in Section IV, we provide a summary and the conclusion of our work.

\section{Theoretical Framework}

The energy associated with a single nanoparticle due to magnetocrystalline anisotropy is given by~\cite{carrey2011,anand2018}
\begin{equation}
E^{}_K=K^{}_{\mathrm{eff}}V\sin^2\theta
\end{equation}
Here $K^{}_{\mathrm {eff}}$ is the uniaxial anisotropy constant, $V$ is the volume of the nanoparticle, and $\theta$ is the angle between anisotropy axis and magnetic moment.
%\subsection{Aggregation and cluster formation in suspensions of magnetic nanoparticles}
%\label{agg}
Clustering is relatively common in suspensions of MNPs or magnetic fluids, as observed in electron microscopy or light scattering experiments~\cite{pei2019,chantrell1982}. 
%Normal magnetic fluids contain $\sim10^{23}$ particles/cc and the collisions between particles are frequent. 
As the particles are magnetised, they adhere and form agglomerates. We can calculate the dipolar interaction energy $E^{}_{\mathrm {dip}}$ of $i^{th}$ nanoparticle in an assembly of MNPs as~\cite{odenbach2002,rosensweig1997,usov2017}
%\begin{equation}
%\label{dipole}
%E_{dd}(s)= -\frac{1}{4\pi \mu_{o}}\left(\frac{3\left(\vec{\mu_{i}}\cdot\vec{s}\right)\left(\vec{\mu_{j}}\cdot \vec{s}\right)}{s^{5}} -\frac{\vec{\mu_{i}}\cdot \vec{\mu_{j}} }{s^3}\right),
%\end{equation}
\begin{equation}
\label{dipole}
E^{}_{\mathrm {dip}}=\frac{\mu^{}_o}{4\pi}\sum_{j,\ j\neq i}\left[ \frac{\vec{\mu_{i}}\cdot\vec{\mu_{j}}}{s^3}-\frac{3\left(\vec{\mu_{i}}\cdot\vec{s}\right)\left(\vec{\mu_{j}}\cdot\vec{s}\right)}{s^{5}}\right].
\end{equation}
Here $\mu_{o}$ is the permeability of free space; $\vec{\mu}_{i}$ and $\vec{\mu}_{j}$ are the magnetic moment vectors of $i^{th}$ and $j^{th}$ nanoparticle respectively, and $s$ is the center-to-center separation between $\mu_{i}$ and $\mu_{j}$. The particle has a magnetic moment $\mu=M^{}_sV$, $M^{}_s$ is the saturation magnetization. 
%$\mu^{}_o\vec{H}^{}_{\mathrm {dip}}$ is the dipolar field due to all other nanoparticles, present in the system. 
The corresponding dipolar field $\mu^{}_o\vec{H}^{}_{\mathrm {dip}}$ is given by~\cite{rosensweig1997,tan2014}
\begin{equation}
\mu^{}_o\vec{H}^{}_{\mathrm {dip}}=\frac{\mu^{}_o}{4\pi}\sum_{j,\ j\neq i}\left[\frac{3\left(\vec{\mu_{j}}\cdot\vec{s}\right)\vec{s}}{s^{5}} -\frac{\vec{\mu_{j}}}{s^3}\right],
\end{equation}  
%Here $\mu_{j}$ is the magnetic moment vector of the $j^{th}$ particle, $s$ is the center-to-center separation between $\mu_{i}$ and $\mu_{j}$; and the $\mu_{o}$ is the permeability of free space.

When the particles are in contact and the moments are aligned, $\vec{\mu_{i}}\cdot \vec{\mu_{j}} = \mu^{2}$ and $\left(\vec{\mu_{i}}\cdot\vec{s}\right)\left(\vec{\mu_{j}}\cdot \vec{s}\right) =  \mu^{2}s^{2}=\mu^{2}D^{2}$. In the latter case, $D$ is the diameter of the nanoparticle. In such a case, the dipole-dipole contact energy $E^{}_{\mathrm {dd}}$ can be evaluated using the following expression~\cite{rosensweig1997}:
%Eq.~(\ref{dipole}) then  reduces to the following form, giving the dipole-dipole contact energy $E^{}_{\mathrm {dd}}$~\cite{rosensweig1997}:
\begin{equation}
\label{c_dd}
E^{}_{\mathrm {dd}} = \frac{1}{12} \mu_{o} M_{s}^{2} V.
\end{equation}
As it is directly proportional to the magnetic volume and $M_{s}$, smaller particles are less likely to aggregate. However, the aggregation can be disrupted by the available thermal energy $k_{B}T^{}_r$, where $k^{}_B$ is the Boltzmann constant and $T^{}_r$ is the temperature. The effectiveness of disruption  is governed by the ratio of the thermal and dipole-dipole contact energy~\cite{rosensweig1997}:
\begin{eqnarray}
\label{ratio}
 E_{R} &=& \frac{k_{B}T^{}_r}{E_{\mathrm {dd}}} = \frac{12k_{B}T^{}_r}{\mu_{o} M_{s}^{2} V}.
\end{eqnarray}
To escape agglomeration, $E^{}_R$ must be greater than unity~\cite{rosensweig1997,odenbach2002} yielding:
\begin{equation}
\label{partsize}
D_{c}^{*} \leq \left(72 k^{}_{B}T^{}_r/ \pi \mu_{o}M_{s}^{2}\right)^{1/3}.
\end{equation}
Thus particles with diameter $D_{c}^{*}$ are on the agglomerating threshold, but those with diameters less than $D_{c}^{*}$ manage to escape this fate. They at the most form small (nano) clusters, e.g., dimers, trimers, tetramers, etc. The mean cluster size is expected to be governed by a balance between the energies responsible for the complementary mechanisms of aggregation and fragmentation. These can be further tailored by choice of surfactant coating and its thickness ~\cite{odenbach2002,eberbeck2006,singh2009,singh2012}.

We apply an oscillating magnetic field to probe the hysteresis behaviour of minimum energy geometrical configurations of $k$-mer. It is given by~\cite{anand2020}
%To study the magnetic hysteresis, we apply an alternating magnetic field ${\mu^{}_{o}H}$ along the $z$-direction (along the chain axis of MNPs) expressed as
\begin{equation}
\mu^{}_{o}H=\mu^{}_oH^{}_{\mathrm {max}}\cos\omega t,
\label{magnetic}
\end{equation} 
where $\mu^{}_{o}H_{\mathrm {max}}$ and $\omega=2\pi\nu$ are the amplitude and angular frequency of the applied magnetic field respectively, $\nu$ is the linear frequency, and $t$ is the time. The total energy of the $i^{th}$ nanoparticle under the influence of dipolar and external magnetic field is given by~\cite{tan2014,anand2019}
\begin{equation}
E^{}_i=K^{}_{\mathrm {eff}}V\sin^2 \theta^{}_i+E^{}_{\mathrm {dip}}-\vec{\mu}^{}_i\cdot\mu^{}_o\vec{H}
\end{equation}
Here $\theta^{}_i$ is the angle between the anisotropy axis and the $i^{th}$ magnetic moment of the system. 
%and $\phi^{}_i$ is the angle between the anisotropy field and the total magnetic field $ H^{}_{\mathrm {total}}$ (dipolar and external field).

Using first-principle calculation, we first identify low energy geometric configurations of the tiny cluster in the absence of an external magnetic field. We then implement kMC simulations to probe the local and the averaged hysteresis response as a function of dipolar interaction strength, particle size and size of $k$-mer. To see the effect of dipolar interaction at the particle level, we also probe the variation of local dipolar field acting on each nanoparticle as a function of the applied magnetic field. In the kMC algorithm, dynamics is captured more accurately, which is essential to study the dynamic hysteresis response of dipolar interacting MNPs~\cite{chantrell2000}.
%study the magnetic hysteresis of dipolar interacting MNPs in a linear chain as a function of the orientation of the anisotropy axis, dipolar interaction strength, frequency, and temperature.  
%In this algorithm, dynamics is more accurately taken into account, which is necessary to study the magnetic hysteresis properties at high frequency. {\color{blue} The upper limit of frequency that the kMC simulations can be used is when the frequency is smaller than the precession of the magnetization time scale,   usually $\sim 10^{-9}$ s.}
Using this technique, we can accurately describe the dynamical properties of MNPs in the superparamagnetic or ferromagnetic regime without any artificial or abrupt separation between them~\cite{tan2014}. We have used the same algorithm, which is described in greater detail in the work of Tan {\it et al.} and Anand {\it et al.}~\cite{tan2014,anand2019}. Therefore, we do not reiterate it here to avoid repetition. The local hysteresis loop area $A^{}_i$ (due to individual nanoparticle) of the $i^{th}$ nanoparticle can be calculated as~\cite{anand2016}
\begin{equation}
A^{}_i=\oint M_i(H)dH,
\label{local_heat}
\end{equation}
The above intergral is evaluated over the entire period of the external magnetic field change. $M^{}_i(H)$ is the magnetization of $i^{th}$ magnetic nanoparticle at magnetic field $H$. 

\section{Simulations Results}
%We consider a one-dimensional chain of the spherically shaped superparamagnetic nanoparticle. We have considered following values of parameters: $D=8$ nm, $M^{}_s=4.77\times10^{5}$ $\mathrm{Am^{-1}}$, $K_{\mathrm {eff}}=13\times10^{3}$ $\mathrm{Jm^{-3}}$, $\nu=10^5$ and $10^9$ Hz. To obtain well-saturated hysteresis curve, the amplitude of applied magnetic field $\mu^{}_{o}H^{}_{\mathrm {max}}$ is taken as 164 mT, which is about three times the anisotropy field $H^{}_K=2K^{}_{\mathrm {eff}}/M^{}_s$ (for $\alpha=0^\circ$). The total number of MNPs is $n=100$ and $\lambda$ has been varied from 0.0 to 1.0. We have performed simulations for a set of temperatures $T=0$, 10, 20, 30, 40, 50, 100, 150, 200, 250 and 300 K which are below and above the blocking temperature $T^{}_B$ that for the material parameters used [$\nu=10^5$ Hz, $\tau^{}_o=0.5\times10^{-10}$ s] is 24.35 K~\cite{bedanta,kolhatkar2013}.

\begin{figure}[!htb]
	\centering\includegraphics[scale=0.47]{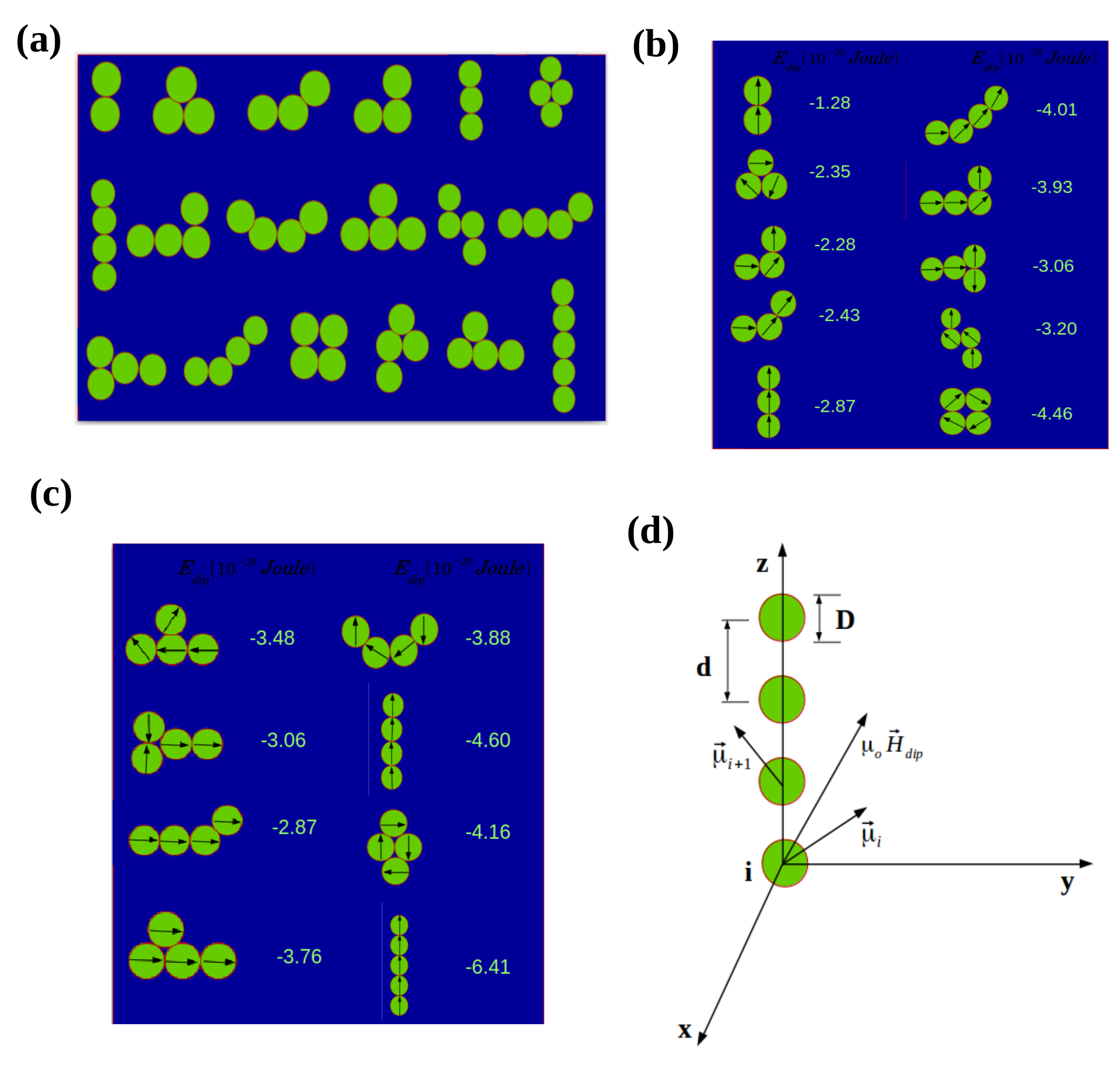}
	\caption{(a) Prototypical geometrical arrangements of $k$-mer: dimer, trimer, tetramer and pentamer considered in our evaluations to obtain low energy configuration. Typical low energy configurations with magnetic moment orientations are shown in (b) and (c). The value of dipolar interaction energy is also depicted for each configuration. It is clearly evident that the chain arrangement of $k$-mer has the lowest interaction energy. (d) Schematic used for kMC simulations to probe the local and averaged hysteresis response; $i$ is the index for particle position in a $k$-mer.}
	\label{figure1}
\end{figure}

%When $r_{c}\lesssim r_{c}^{*}$, the toxicity limits allow the magnetite particle concentration to be in the range of $10^{14}$ to $10^{16}$ particles/cm$^{3}$~\cite{singh2012}. 
We can expect clusters of fewer particles due to the ubiquitous dipole-dipole interactions in MNPs carriers. These tiny clusters are expected to have significantly different hysteresis behaviour from the monomers and to see this; it is essential to identify the cluster geometries and their spin configurations. Although this task is humongous, the following simplifications make our analysis tractable. We choose representative geometric arrangements for dimers, trimers and tetramers, pentamers, etc. in two dimensions. We do not expect a significant loss of information on this account as the clusters are tiny. These prototypical arrangements are depicted in Fig.~\ref{figure1}(a). Each cluster geometry is then decorated by magnetic moments. They are allowed to assume nine orientations, specified by angles $n\pi/4$, $n = 0, 1,2, 3..,8$. Thus, if $G$ is the number of geometric configurations for a $k$-mer, the total configurations are $G\times 9^{k}$. For instance, a trimer could be formed in $4\times 9^{3} = 2916$ ways and a tetramer in $12\times 9^{4} = 78732$. So the number of configurations in our simplified model is large even for tiny clusters and grows exponentially with cluster size. The particle is assumed to have a diameter $D=8$ nm to perform model calculations. The other material parameters are: $K^{}_{\mathrm {eff}}=13\times10^{3}$ $\mathrm{Jm^{-3}}$ and $M^{}_s=4.77\times10^{5}$ $\mathrm{Am^{-1}}$. These parameters correspond to magnetite (Fe$^{}_3$O$^{}_4$), one of the best candidates for biomedical applications due to its biocompatibility~\cite{fuentes2018}. As discussed above, we have used Eq.~(\ref{dipole}) to evaluate all the possible configurations's dipolar interaction energy. At body temperature $T^{}_b\approx310$ K, the thermal energy $ k_{B}T_{b} \approx 4.29\times10^{-21}$ Joule. So, clusters with energies $\sim O(k^{}_{B}T^{}_r)$ will be unstable. Therefore, we consider only the minimum energy configurations, which are not destroyed by the thermal energy. We have shown typical morphologies of the tiny cluster with their dipolar interaction energy in Fig.~\ref{figure1}(b) and Fig.~\ref{figure1}(c). It is evident that magnetic moments tend to form head to the tail arrangement as the corresponding structure's dipolar interaction energy is the minimum compared to the other arrangements. One of the most interesting facts to note is that the chain arrangement of MNPs is the most stable of all the possible geometrical structure. This observation is in perfect agreement with the earlier works~\cite{orue2018,morales2018,serantes2014}. Researchers have also found the chain-like arrangement of $k$-mers in liposome~\cite{bealle2012,alonso2016}.
%So, the first principle calculations reveals chain arrangment of MNPs is more favourable.}

From the above observations, it is evident that the chain arrangement of $k$-mer is an ideal candidate to probe the fundamental effects of dipolar interactions on local and averaged hysteresis response in the presence of an external alternating magnetic field. Therefore, in the rest of the paper, we study the physical processes involved in local (due to individual nanoparticle) and global hysteresis mechanism under the influence of dipolar interaction in the chain-like structure of $k$-mer. Our study is primarily focused on gathering and understanding the essential effects caused by dipolar interactions in magnetic hyperthermia experiments. The anisotropy axes of the MNPs are assumed to have random orientations to remain closer to the real experiments. To vary the dipolar interaction, we define a scale parameter $\lambda=D/d$, where $d$ is the centre to centre distance between two consecutive MNPs in a $k$-mer as shown in the schematic Fig.~\ref{figure1}(d). So, $\lambda=0.0$ signifies non-interacting case while $\lambda=1.0$ corresponds to strongly interacting MNPs. All the numerical simulations presented in this work are performed at $T^{}_r=300$ K and $\nu=10^5$ Hz, and the external magnetic field is applied along the $z$-direction [please see  Fig.~\ref{figure1}(d)]. The applied magnetic field's amplitude is taken as $\mu^{}_oH_{\mathrm {max}}=0.06$ T, which is slightly larger than single-particle anisotropy field $H^{}_K=2K^{}_{\mathrm {eff}}/M^{}_s$~\cite{carrey2011}.

\begin{figure}[!htb]
	\centering\includegraphics[scale=0.47]{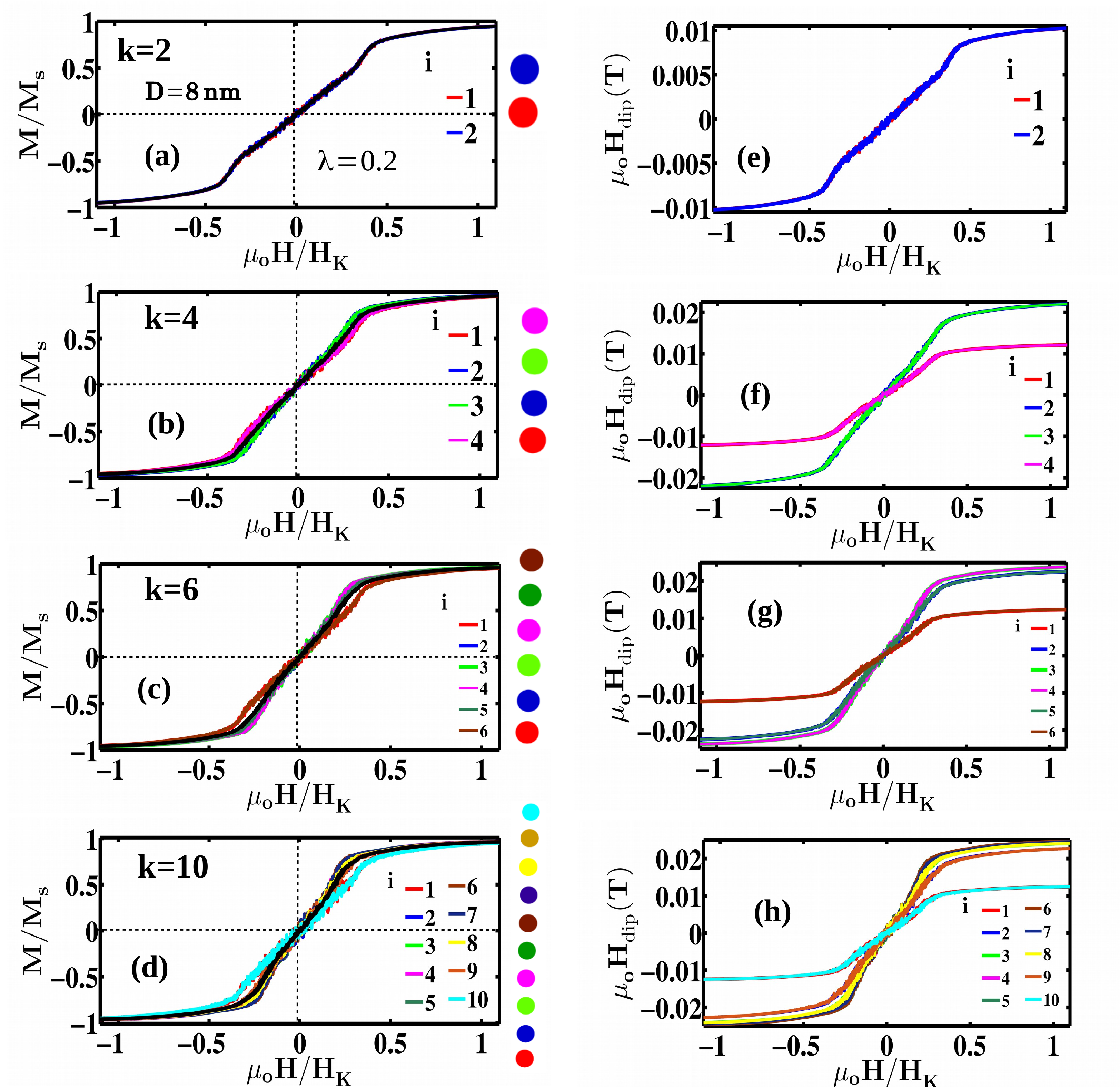}
	\caption{The local and averaged hysteresis curve is shown for various $k$-mer. The variation of correspoding local dipolar field as a function of an applied magnetic field is also shown. We have considered four values of $k$-mer: $k=2$ [(a) and (e)], $k=4$ [(b) and (f)], $k=6$ [(c) and (g)], and $k=10$ [(d) and (h)]. Local hysteresis curve and dipolar field are shown with the same colour; the averaged hysteresis curve is shown with a black line, $i$ is the particle position in a $k$-mer. %The local hysteresis loop are shown with different colour to distinguish from each other and indexed using $i$, where $i$ is the particle position in the k-mer. We have also shown the $k$-mer for each considered value of $k$ and pariicles are coloured with same color as that of $i$.
The particle size is $D=8$ nm and dipolar interaction strength $\lambda=0.2$. The local dipolar field's variation is exactly similar as that of local hysteresis response. The coercive field and remanence magnetization have negligible values, indicating the dominance of superparamagnetic character. 
}
\label{figure2}
\end{figure}

In Fig.~(\ref{figure2}) we plot the local (individual nanoparticle), and the averaged magnetic hysteresis for four values of $k$-mers ($k=2$, 4, 6 and 10). To probe the effect of dipolar interaction at the particle level, we also analyze the variation of local dipolar field acted on each nanoparticle as a function of the external magnetic field. The particle size is taken as $D=8$ nm and dipolar interaction strength $\lambda=0.2$. The local hysteresis curve and dipolar field variation are shown with different colour depending on the nanoparticle's position in a particular $k$-mer. The averaged hysteresis curve is shown with black colour.
%We have chosen particle size $D=8$ nm, very weak dipolar interaction strength $\lambda=0.2$. 
As the particle size lies in the superparamagnetism regime, and dipolar interaction is negligibly small, the local and averaged hysteresis loop show negligibly small value of the coercive field and remanent magnetization. The local hysteresis depends strongly on the corresponding dipolar field acting on a nanoparticle. For instance, the dipolar field acted on each nanoparticle is the same in the dimer ($k=2$). As a result, the local hysteresis curve is exactly the same for both particles. Similarly, in the case of tetramer ($k=4$), the dipolar field acted on nanoparticles at both the ends ($i=1$ and $i=4$) is the same. In contrast, the remaining two nanoparticles experience same dipolar field. Consequently, the local hysteresis curve for nanoparticles positioned at $i=1$ and $i=4$ is exactly similar to each other; the remaining MNPs have identical hysteresis curves. We can infer similar observation for hexamer ($k=6$) and decamer ($k=10$). These observations are in qualitative agreement with the theoretical works of Torche {\it et al.}~\cite{torche2020}. It is also in qualitative agreement with the works of  Vald\'es {\it et al.}~\cite{valdes2020}. The averaged hysteresis curve is found to be the mean of local hysteresis curves of constituent MNPs. Interestingly, the variation of the local dipolar field acted on individual nanoparticle is the same as that of corresponding magnetic hysteresis response. As the size of $k$-mer increases, there is more uniformity in constituent nanoparticles's local
hysteresis behaviour except for MNPs at the both ends.

\begin{figure}[!htb]
	\centering\includegraphics[scale=0.47]{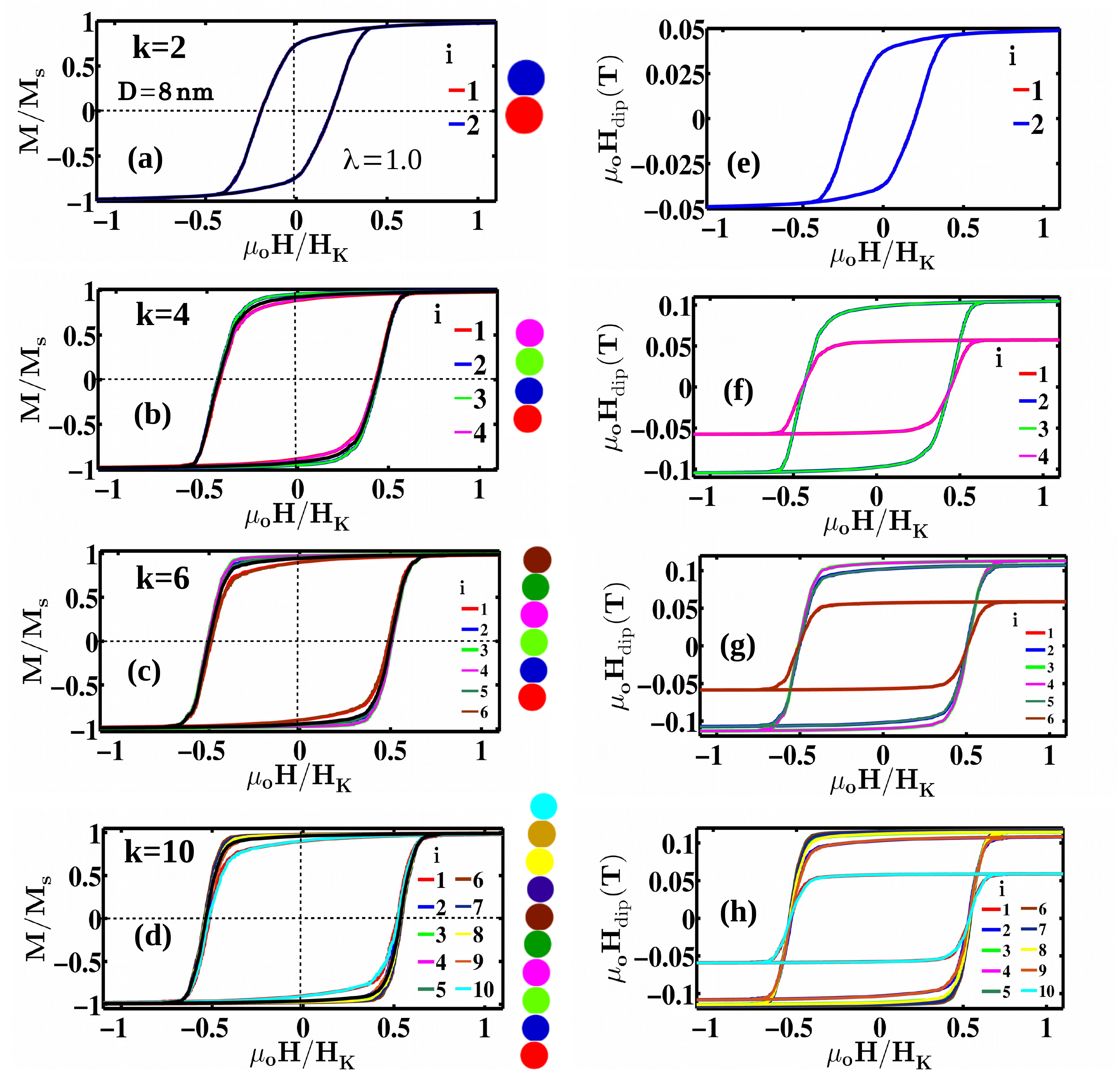}
	\caption{ The study of local and the averaged hysteresis curve for large dipolar interaction strength $\lambda=1.0$ [(a)-(d)]. We have also shown the corresponding  variation of local dipolar field acted on each particle in a $k$-mer [(e)-(f)]. The particle size is taken as $D=8$ nm. Even in the case of sizeable dipolar interaction, the local hysteresis curve is intimately related to the amount of dipolar field acted on it. There is an increase in averaged and local hysteresis loop area with an increase in the size of $k$-mer.}
	\label{figure3}
\end{figure}

Next, we study the hysteresis behaviour for strongly dipolar interacting MNPs. In Fig.~(\ref{figure3}), we plot the local and averaged hysteresis curves for $\lambda=1.0$. All the other parameters are the same as that of Fig.~(\ref{figure2}). Even in the case of large dipolar interaction strength, the shape of the local hysteresis curve for both the constituent particles is identical for dimer structure. It is due to the fact that the same dipolar field is acting on each particle in the dimer ($k=2$). The value of the dipolar field experienced by the nanoparticle is way larger than the previous case. The variation of local dipolar field acted on each nanoparticle is the same as that of corresponding local hysteresis response irrespective of the size of $k$-mer. For all the $k$-mer, the local hysteresis response of the constituting nanoparticles is dictated by the dipolar field acted on it. For $k>10$, the local hysteresis curves are uniform except for MNPs at both the ends (curve are not shown).  The value of the coercive field and remanent magnetization are also very large irrespective of the size of the $k$-mer. It can be attributed to the enhanced ferromagnetic coupling between the MNPs. The area under the hysteresis curve is also huge as compared to the weakly interacting case.

\begin{figure}[!htb]
	\centering\includegraphics[scale=0.47]{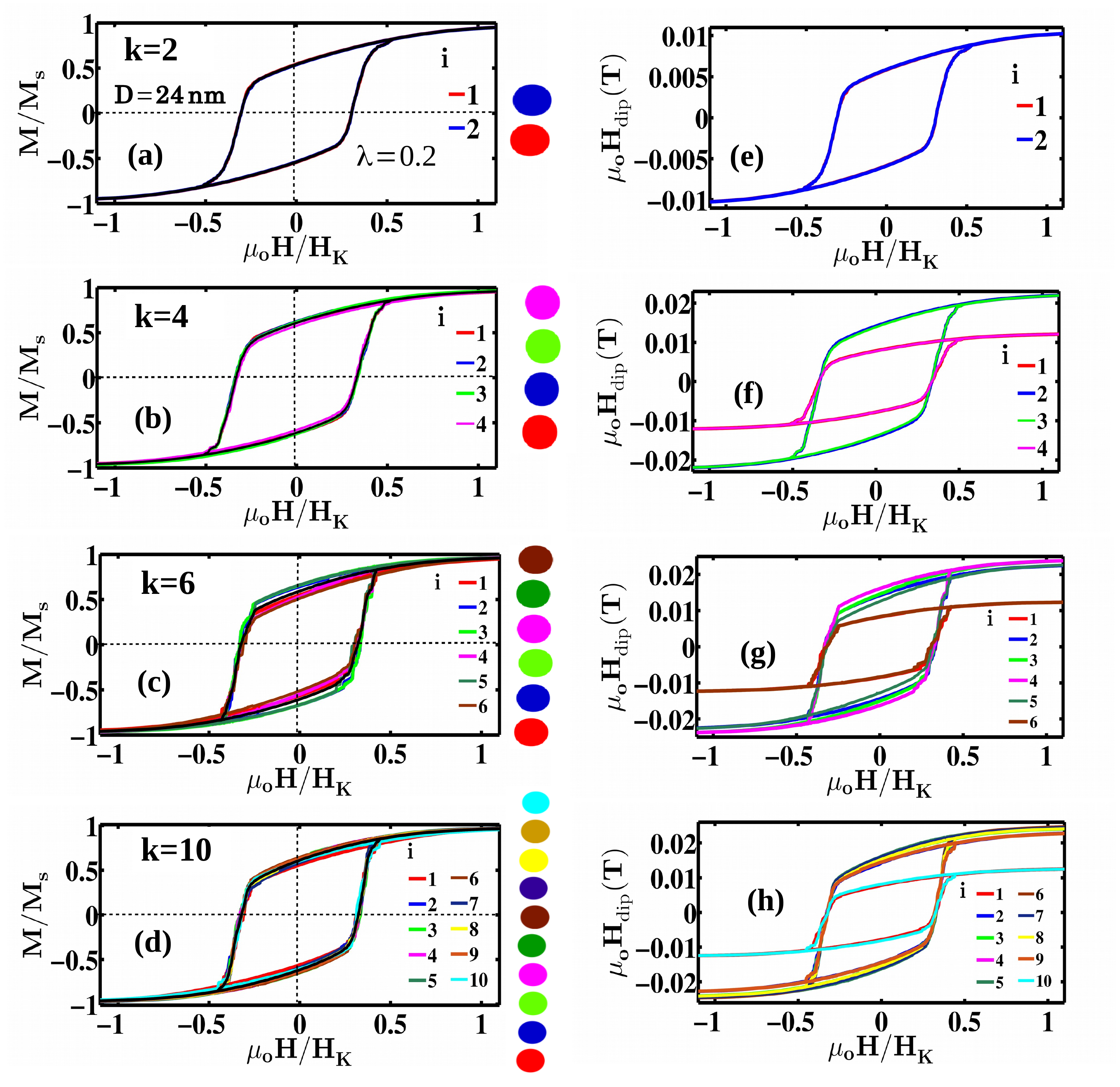}
	\caption{The local and averaged hysteresis response variation for ferromagnetic nanoparticle ($D=24$ nm) and weak dipolar interaction ($\lambda=0.2$). We have also shown the local dipolar field's variation as a function of an external magnetic field. We have considered four values $k$-mer: $k=2$ [(a) and (e)], $k=4$ [(b) and (f)], $k=6$ [(c) and (g)], and $k=10$ [(d) and (h)]. It is clearly seen that the local and the averaged hysteresis loop area is huge compared to $D=8$ nm, even in the case of weak dipolar interaction. The local hysteresis is also directly proportional to the corresponding dipolar field acted on it.} 
	\label{figure4}
\end{figure}

\begin{figure}[!htb]
	\centering\includegraphics[scale=0.47]{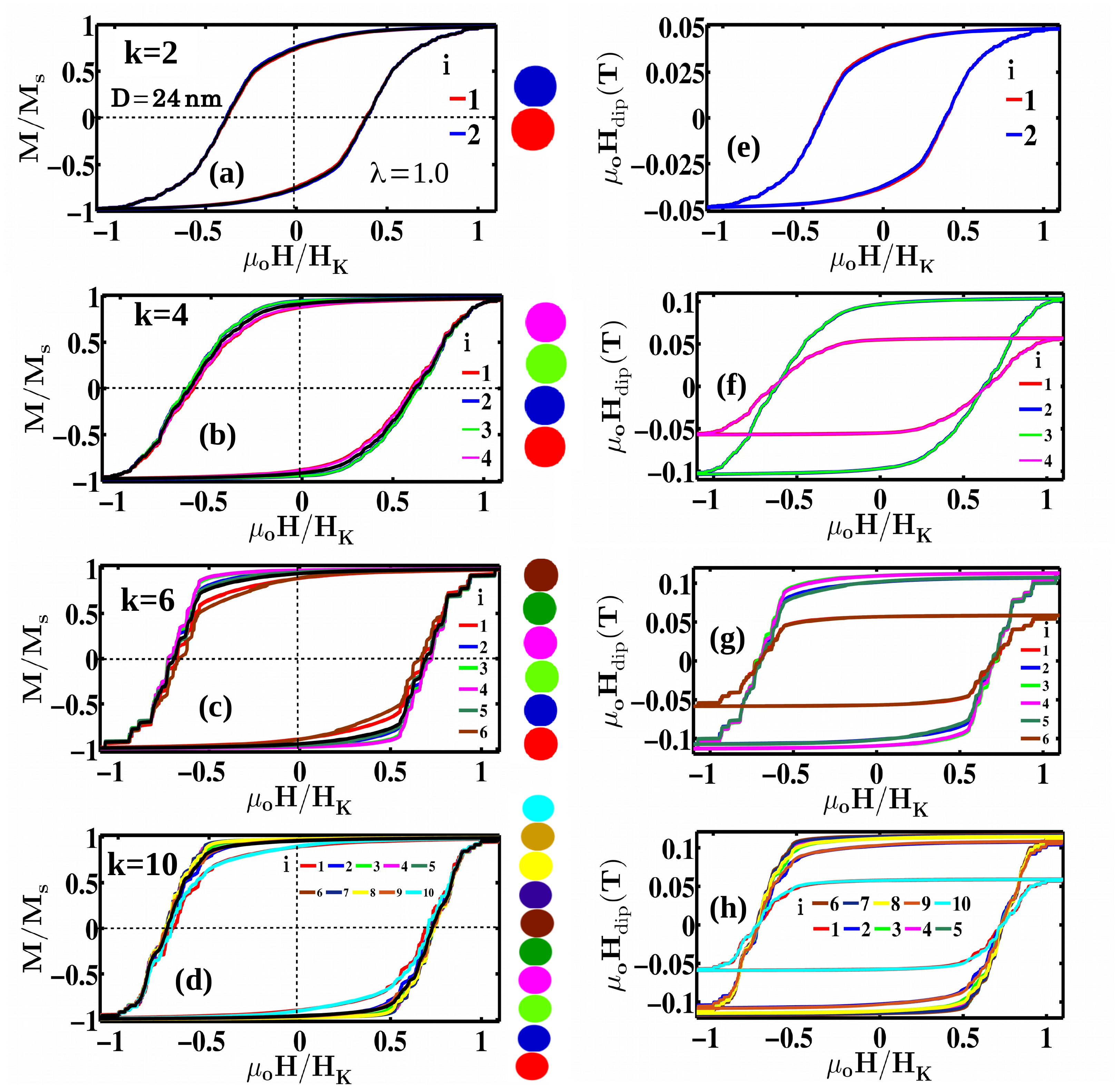}
	\caption{The local and averaged hysteresis curve are shown for strongly interacting MNPs ($\lambda=1.0$), particle size $D=24$ nm, and four values of $k$-mer ($k=2$, 4, 6, and 10) [(a)-(d)]. The corresponding variation of the local dipolar field as a function of an applied magnetic field is shown in (e)-(h). There is an increase in the hysteresis loop area and dipolar field with the size of $k$-mer. Irrespective of the size of $k$-mer, the variation of the local dipolar field is the same as that of the corresponding hysteresis response.}  
	\label{figure5}
\end{figure}

What happens to the averaged and local hysteresis response in the case of the ferromagnetic nanoparticle? To analyze it, we study the magnetic hysteresis and variation of local dipolar field acted on each nanoparticle in a $k$-mer as a function of an external magnetic field for $D=24$ nm with weak dipolar interaction $\lambda=0.2$ [Fig.~(\ref{figure4})] and strongly dipolar interacting ($\lambda=1.0$) [Fig.~(\ref{figure5})]. Even with ferromagnetic nanoparticles, the local and global hysteresis response is dictated by the dipolar field acted on the constituent MNPs. The variation of the local dipolar field acted on each nanoparticle is exactly similar to that of the corresponding hysteresis curve, irrespective of the size of the $k$-mer and dipolar interaction strength. Interestingly, the shape of the hysteresis curve is like a perfect square for strongly interacting MNPs even with randomly oriented anisotropy. It means that dipolar interaction creates an additional anisotropy so-called shape anisotropy which weakens the impact of randomly oriented uniaxial anisotropy. Consequently, remanence also gets enhanced with an increase in the size of $k$-mer and dipolar interaction strength.  The coercive field and remanent magnetization have larger values even in the case of weak dipolar interaction. The hysteresis has less dependence on the size of $k$-mer for weak dipolar interacting MNPs. The hysteresis loop area is very large compared to superparamagnetic nanoparticle even in the case of the weak interacting case [please see Fig.~(\ref{figure4})]. On the other hand, there is an increase in the area under the hysteresis curve with an increase in the size of the $k$-mer for strongly interacting MNPs. These results could help choose precise values of particle size, and interaction strength to optimize the heat dissipation for drug delivery and hyperthermia applications.  
%Interestinly, there is also an increase in remanence with $k$, which has intimate relation with dipolar interaction as the latter is also increasing with the size of $k$-mer. {\color{blue} try to link the remancence with shape anisotropy. The remanence magnetization is increasing as the size of the $k$-mer grows. It has intimate relatation with the dipolar field as it is clearly evident that magntitude of dipolar is also increasing with an increase in the size of $k$-mer at zero external magnetic field ($\mu^{}_oH/H^{}_K=0$) [Please see Fig.~\ref{figure5} (e)-(h)].}

\begin{figure}[!htb]
	\centering\includegraphics[scale=0.30]{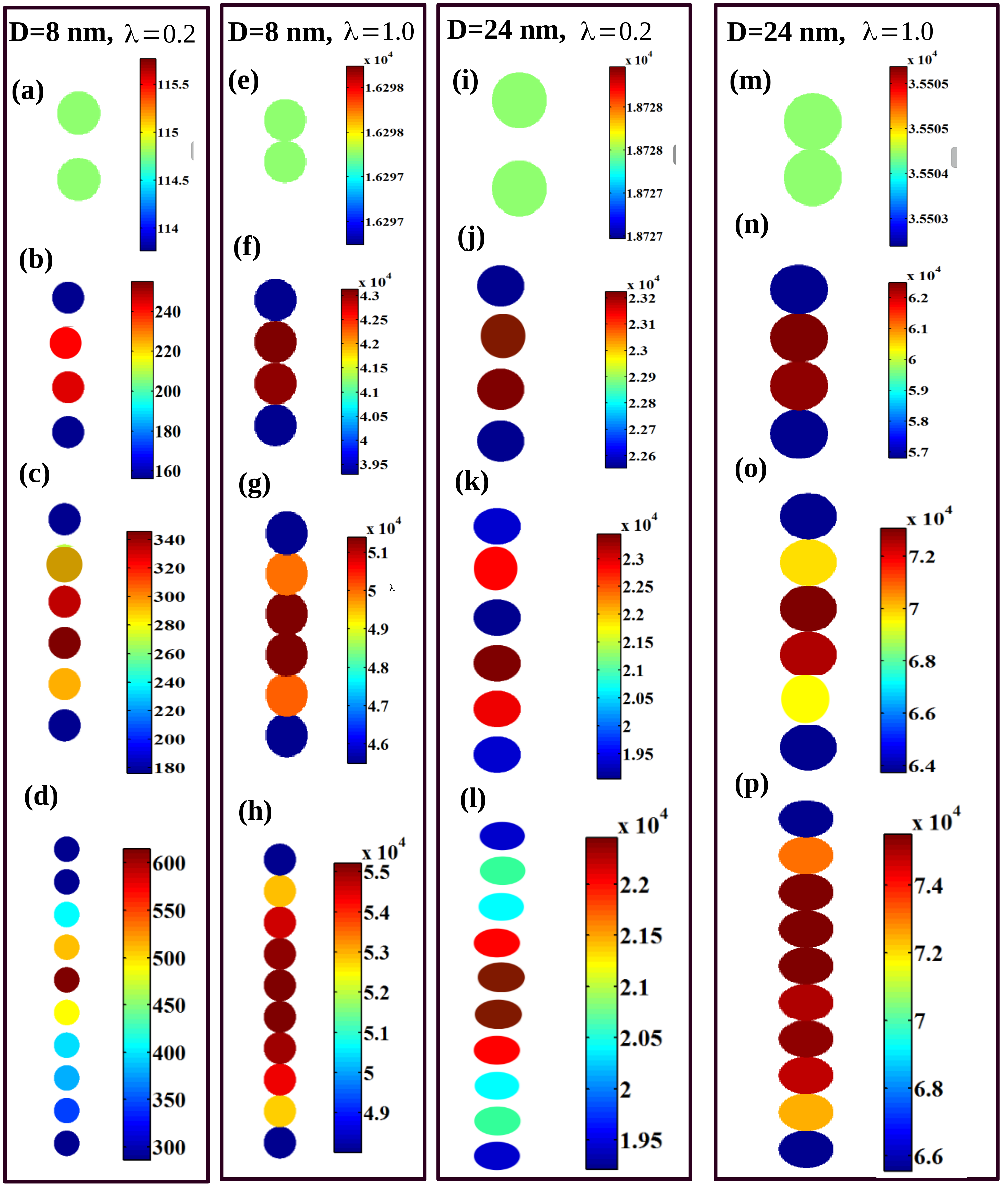}
	\caption{We have depicted the value of the local hysteresis loop area $A^{}_i$ (due to individual nanoparticle) for four values of $k$-mer: $k=2$, 4, 6 and 10. The other parameters are: $D= 8$ nm, $\lambda=0.2$ [(a)-(d)]; $D=8$ nm, $\lambda=1.0$ [(e)-(h)]; $D=24$ nm, $\lambda=0.2$ [(i)-(l)]; and $D= 24$ nm, $\lambda=1.0$ [(m)-(p)]. There is a wide distribution of $A^{}_i$ in smaller $k$-mer, as the local dipolar field varied rapidly with position of nanoparticle in a $k$-mer. On the other hand, there is more uniformity in $A^{}_i$ for the larger size of $k$-mer. There is an increase in $A^{}_i$ with the size of $k$-mer, particle size and dipolar interaction strength.} 
	\label{figure6}
\end{figure}

To quantify the above observations, we analyze the local hysteresis loop area $A^{}_i$ of constituent MNPs for four values of $k$-mer ($k=2,4,6$, and 10) with particle size $D=8$ and 24 nm. We have considered two representative values of dipolar interaction strength ($\lambda=0.2$) [Fig.~\ref{figure6}(a)-\ref{figure6}(d) and Fig.~\ref{figure6}(i)-\ref{figure6}(l)] and $\lambda=1.0$ [Fig.~\ref{figure6}(e)-\ref{figure6}(h) and Fig.~\ref{figure6}(m)-\ref{figure6}(p)]. The legend is also shown for each case. It is evident that $A^{}_i$ has an intimate relationship with local dipolar field acting on a nanoparticle. For instance, as the dipolar field acted on both particles in the dimer ($k=2$) is equal, the corresponding local hysteresis loop area $A^{}_i$ is also equal to each other. We can draw similar observations for other values of $k$-mer, irrespective of dipolar interaction strength and particle size. The value of $A^{}_i$ is larger for strongly interacting MNPs as compared to the weakly dipolar interacting case. The same is true for ferromagnetic nanoparticle $D=24$ nm even for weak interacting MNPs. It is clear that $A^{}_i$ of the nanoparticles at the boundary is smaller than the central MNPs. It can be explained from the fact that 
the boundary MNPs experience a smaller dipolar field than the MNPs at the centre.
%As a consequence, the hysteresis loop area of the MNPs at the boundary is smaller as compared to the cetrnal MNPs. 
For smaller $k$-mer, there is a large variation in $A^{}_i$ as local dipolar field acted on the nanoparticle in such a case varies rapidly as a function of the position of MNPs. On the other hand, the local dipolar field acted on the particle in larger $k$-mer has less variation except for MNPs at the boundary. Therefore, there is more uniformity in $A^{}_i$ for larger $k$-mer. 
%For $k>2$, $A^{}_i$ for the centrally positioned nanoparticle is the maximum and it decreases as .
$A^{}_i$ is directly proportional to the dipolar field, acting on the individual particle. The larger the dipolar field, the larger the hysteresis loop area. The local hysteresis loop area $A^{}_i$ increases with an increase in the size of $k$-mer. It also increases with an increase in dipolar interaction strength and particle size. These observations can help tune the dipolar interaction to optimize the local and averaged hysteresis response, which is essential for hyperthermia applications.
 
\begin{figure}[!htb]
	\centering\includegraphics[scale=0.47]{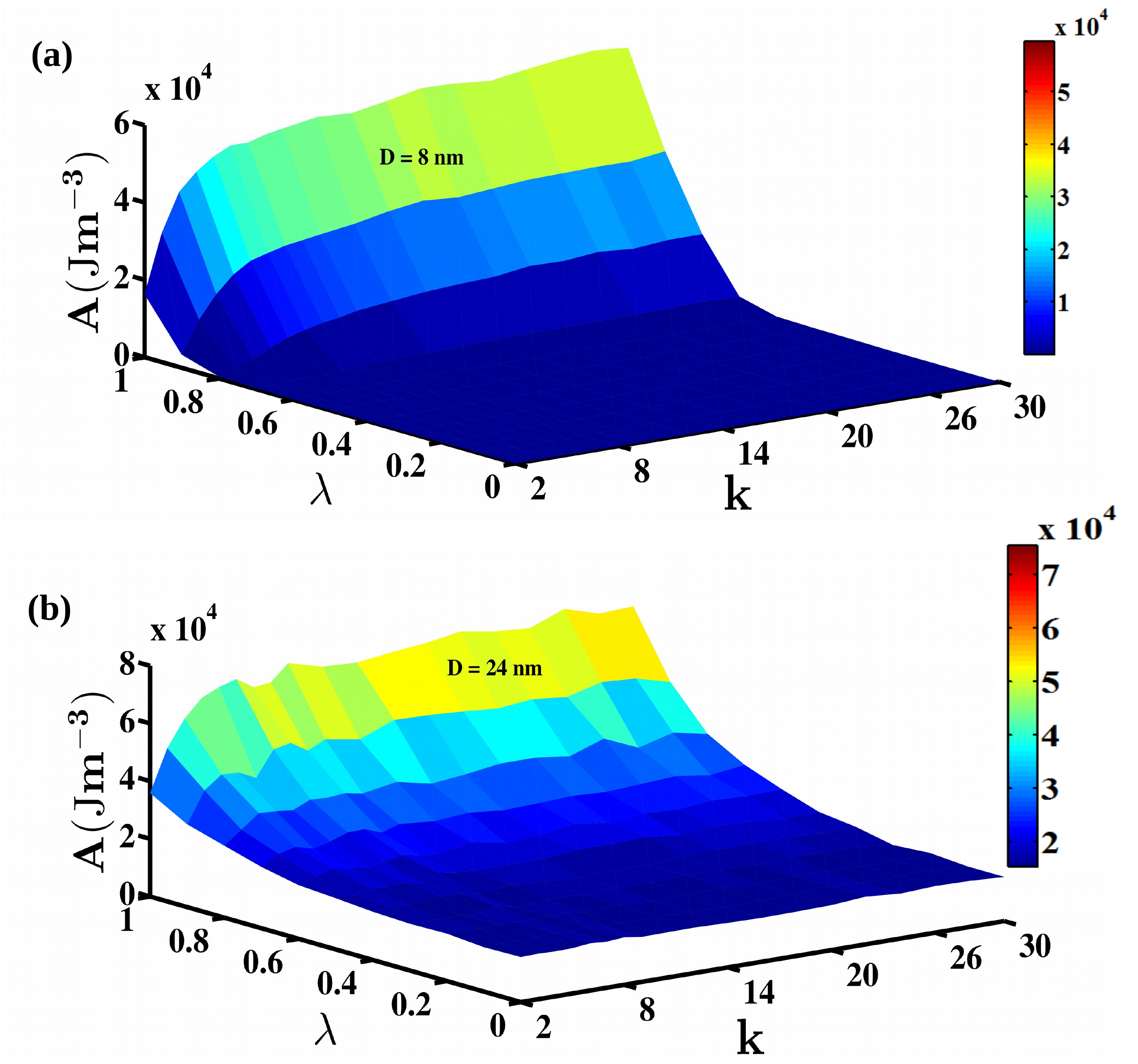}
	\caption{The variation of averaged hysteresis loop area $A$ (due to entire system) as a function of dipolar interaction strength $\lambda$ and size of $k$-mer for $D=8$ nm [(a)] and $D=24$ nm [(b)]. For superparamagnetic nanoparticle ($D=8$ nm) and weak dipolar interaction strength ($\lambda\leq0.6$), $A$ is significantly small and depends weakly on $k$. While for ferromagnetic nanoparticle, $A$ is more significant even with weakly interacting MNPs. There is an increase in $A$ with $k$ and $D$. Irrespective of particle size, the value of $A$ saturates for $k\geq20$ and considerable dipolar interaction strength.}
	\label{figure7}
\end{figure}

Finally, we study the dependence of the averaged hysteresis loop area $A$ on the size of $k$-mer and dipolar interaction strength. In Fig.~(\ref{figure7}), we plot $A$ as a function of $k$ and $\lambda$ for $D=8$ and 24 nm. $k$ is changed from 2 to 30, and $\lambda$ is varied from 0 to 1.0. In the case of superparamagnetic nanoparticle ($D=8$ nm) and small dipolar interaction strength ($\lambda\leq0.6$), $A$ is very small and depends weakly on the size of $k$-mer. On the other hand, there is an increase in $A$ with $k$ for strongly interacting MNPs ($\lambda>0.6$). 
%There is an increase in the value of $A$ with $k$ for large dipolar interaction strength $\lambda>0.6$. $A$ also increases with $\lambda$ irrespective of $k$.
Interestingly, the value of $A$ is more considerable for ferromagnetic nanoparticle ($D=24$ nm) as compared to superparamagnetic counterpart ($D=8$ nm) even for weakly dipolar interacting MNPs ($\lambda\leq 0.6$). $A$ is significantly large for $D=24$ nm in comparison with $D=8$ nm, irrespective of $\lambda$ and $k$. Interestingly, the value of $A$ saturates for $k>20$, irrespective of $D$. It is in perfect agreement with the recent work of Vald\'es {\it et al.} \cite{valdes2020}. These findings could be useful for optimizing the size of $k$-mer and interaction strength to control the heat dissipation, which is essential for hyperthermia applications. 
%Interestingly, $A$ saturates for $k>20$, irrespective of the value of $D$. 
%We consider an ideal chain of low-anisotropy NPs with their easy axes of effective uniaxial anisotropy in the direction of the chain, allowing different orientations of the chain with respect to the applied field.

%A major contributing factor to the slow advances in MNH is the lack of understanding of the physical processes underlying the heating mechanism.
	
%From the aforementioned discussion, ideal NP-chain arrangements are good systems for the study and characterization of the fundamental effects of dipolar interactions on the magnetic response under an ac magnetic field. In this paper, we study in detail the mechanisms involving dipolar interactions in the magnetic relaxation of an ideal system for MFH experiments. 

%Our system is built up in a simple way to gather and understand the essential effects caused by interactions in these experiments. We consider an ideal chain of low-anisotropy NPs with their easy axes of effective uniaxial anisotropy in the direction of the chain, allowing different orientations of the chain with respect to the applied field. The low-anisotropy assumption means that the anisotropy field HK is lower than the amplitude of the ac applied field H 0 . For this system, HK<H0 and interactions are considered; consequently the linear response theory [30] cannot be applied. Hence, another nonlinear model [13,31] is used for this study.}
%\newpage
\section{Summary and Conclusion}
Now, we summarize and discuss the main results presented in this work. Using first-principles calculations, we first identified minimum energy configurations of clusters of $k$ MNPs or $k$-mers (k = 2, 3, 4, 5, etc.). After that, we have used kinetic Monte Carlo simulations technique to probe the local (individual nanoparticle) and averaged hysteresis response as a function of cluster size, dipolar interaction strength and particle size. To investigate the effect of dipolar interaction at the particle level, we also analyze the variation of the local dipolar field acted on each nanoparticle as a function of the applied magnetic field. Dipolar interaction is found to be the primary factor which helps in clustering of MNPs. One can confine the agglomeration process to fewer MNPs by particle size selection. Even for these tiny clusters, there is a possibility of a huge number of configurational arrangments. Of all the possible configurations, chains of $k$-mers are most stable as they have the lowest interaction energy. 
%Dipolar interactions cause clustering in MFs which when uncontrolled creates large aggregates and long chains. Aggregation can be confined to two to four MNPs by size-selection. Though tiny, these {\it nano-clusters} have an exponentially large number of configurational arrangements.The lowest energy configurations are chains with magnetic moments aligned head-to-tail. By size-selection, it is possible to confine the length to two to four MNPs. A good option is to choose particles with core radius $r_{c} \lesssim r_{c}^{*}$ and induce nano-chains by the application of a (small) magnetic field.
The local hysteresis behaviour is primarily dictated by the dipolar field acting on a particle. The variation of the local dipolar field acting on a nanoparticle is precisely similar to that of the corresponding hysteresis curve. For smaller $k$-mer, there is a considerable variation in the local dipolar field as a function of nanoparticle's position in the $k$-mer. Consequently, the local hysteresis response also varies rapidly. For superparamagnetic nanoparticle and weak dipolar interaction, the hysteresis shows zero coercivity and remanence. While for sizeable dipolar interaction strength, the coercive field and remanent magnetization have significant values. The same is true for ferromagnetic nanoparticle with negligible dipolar interaction. Interestingly, the shape of the hysteresis curve is like a perfect square with large dipolar interaction strength.

The local hysteresis loop area $A^{}_i$ of a nanoparticle is  directly proportional to the corresponding local dipolar field. In the case of superparamagnetic nanoparticle and weak dipolar interaction, $A^{}_i$ is small, irrespective of the size of $k$-mer.
%The local hysteresis loop area is very small for superparamgnetic nanoparticle and small dipolar interaction strength.
On the hand, $A^{}_i$ is very large for strongly interacting nanoparticles because of enhanced ferromagnetic coupling.
%the local hysteresis loop area has very large value for strongly interacting MNPs. 
The same is true for ferromagnetic nanoparticle with small dipolar interaction strength. In a $k$-mer $(k>2)$, $A^{}_i$ is the maximum for centrally positioned nanoparticle and its value decrease as we move towards both ends of the $k$-mer. 
%There is large variation in the local hysteresis loop area for smaller $k$-mer.
There is more uniformity in $A^{}_i$ for larger $k$-mer except for MNPs at  both the ends. Similarly, there is a strong dependence of averaged hysteresis loop area $A$ on the dipolar interaction strength. $A$ is very small for small values of dipolar interaction strength, irrespective of $k$-mer. There is an increase in $A$ with the size of $k$-mer and $\lambda$. In the case of ferromagnetic nanoparticle, $A$ is very large even with small dipolar interaction strength. Therefore, it is evident that the local and averaged hysteresis loop area increases with an increase in dipolar interaction strength. Irrespective of particle size and dipolar interaction strength, the value of $A$ saturates for large size of $k$-mer ($k>20$).   

In conclusion, we have studied the local and averaged hysteresis response in the chain-like tiny clusters. We have also probed the local dipolar field at the particle level as a function of an external magnetic field. Our results suggest that the local hysteresis is dictated by the dipolar field acted on a particle. There is also a significant variation in the local hysteresis loop area as a function of  nanoparticle position in a $k$-mer. We emphasize that our methodologies are generic and applicable to a various tiny cluster in distinct physical settings. Our methodologies could provide a theoretical basis to the often used ad-hoc procedures in therapeutic applications. We believe the similar observations can be drawn for the diverse clusters as long as the magnetic behaviour is dictated by the dipolar interaction.
\section*{DATA AVAILABILITY}
The data that support the findings of this study are available from the corresponding author upon reasonable request.
\bibliography{ref}
\newpage

\newpage

\newpage

\newpage

\newpage

\newpage

\newpage
\end{document}